# Handedness manipulation of propagating antiferromagnetic magnons

Yoichi Shiota[1,2,*], Tomohiro Taniguchi[3], Daiju Hayashi[1], Hideki Narita[1], Shutaro Karube[1,2], Ryusuke Hisatomi[1,2], Takahiro Moriyama[4], and Teruo Ono[1,2]

[1] *Institute for Chemical Research, Kyoto University, Uji, Kyoto 611-0011, Japan*
[2] *Center for Spintronics Research Network, Institute for Chemical Research, Kyoto University, Uji, Kyoto 611-0011, Japan*
[3] *National Institute of Advanced Industrial Science and Technology (AIST), Research Center for Emerging Computing Technologies, Tsukuba, Ibaraki 305-8563, Japan*
[4] *Department of Materials Physics, Nagoya University, Nagoya, Aichi 464-8603, Japan*
*e-mail: shiota-y@scl.kyoto-u.ac.jp

**Antiferromagnetic magnons possess a distinctive feature absent in their ferromagnetic counterparts: the presence of two distinct handedness modes, the right-handed (RH) and left-handed (LH) precession modes[1–3]. The magnon handedness determines the sign of spin polarization carried by the propagating magnon, which is indispensable for harnessing the diverse functionalities[4–9]. However, the control of coherently propagating magnon handedness in antiferromagnets has remained elusive so far. Here we demonstrate the manipulation and electrical readout of propagating magnon handedness in perpendicularly magnetized synthetic antiferromagnets (SAF). We find that the antiferromagnetic magnon handedness can be directly identified by measuring the inverse spin Hall effect (ISHE) voltage, which arises from the spin pumping effect caused by the propagating antiferromagnetic magnons in the SAF structure. The RH and LH modes of the magnon can be distinguishable particularly when the SAF structure is sandwiched by heavy metals with the same sign of spin Hall angle. Moreover, we succeed in controlling the handedness of propagating antiferromagnetic magnons by tuning the excitation microwave frequency. This work unveils promising avenues for harnessing magnon unique properties in antiferromagnet-based magnonic applications.**

The control of magnetization by spin current, which is a flow of spin angular momentum, plays an important role in both basic and applied point of view[10,11]. In magnetic materials, spin waves or magnons are known to be the carriers of spin currents. Magnon spin currents are attracting attention as a propagation method with less dissipation than conduction electron spin currents[12–17], because magnon propagation does not involve charge transfer resulting in the absence of Joule heating. Therefore, magnons are expected to be a promising platform for realizing low-power devices, logic circuits, and computing applications[18–29]. In contrast to ferromagnets (FMs), collinear antiferromagnets possess two precession modes, right-handed (RH) precession and left-handed (LH) precession[1–3]. This allows to manipulate the magnon polarization degree of freedom, which is analogous to its optical counterpart. The control of the magnon handedness is a critical issue in applications using magnon polarization degree of freedom, such as magnon field-effect transistor[4,5], domain wall as magnon polarizer and retarder[6], bidirectional domain wall motion[7,8], and quantum system with magnon[9]. Therefore, manipulation and electrical detection of the propagating magnon handedness is highly desirable for practical applications. However, experimental progress is largely limited by the difficulty of controlling antiferromagnetic spin dynamics due to their field-immunity and ultrafast spin dynamics.

In this study, we report the first demonstration of polarization-resolved detection and handedness manipulation of propagating antiferromagnetic magnons. We prepared the synthetic antiferromagnet (SAF) sandwiched by heavy metal (HM) layers. Propagating magnons in SAF structure are excited by applying microwave current to a stripline antenna and emit handedness-dependent spin current via spin pumping effect into HM layers. The spin current is converted to electric current through the inverse spin Hall effect (ISHE)[30]. We find that the RH and LH modes of the magnon provide two distinguishable resonance peaks in the ISHE voltage, particularly when the sign of spin Hall angle in two HM layers is the same; therefore, the magnon handedness can be electrically detected. We further demonstrate the manipulation of magnon handedness by tuning the excitation frequency under conditions where the degeneracy of two magnon modes is lifted. These results lead to a new path for utilizing antiferromagnetic magnons in future polarization-based magnonic devices.

## Results
### Concept of magnon excitation and detection

Our proposal for the excitation and detection of the handedness-dependent magnon is based on the fact that two (RH and LH) precession modes in colinear antiferromagnets can have different resonance conditions, as shown in Fig. 1a. There are two distinct modes corresponding to RH and LH precession of the antiferromagnetic resonance, except at a certain magnetic field where they are degenerated. The opposite linear dependence of the two modes on the magnetic field arises from the fact that the angular momenta of the two modes point in opposite directions, resulting in different Zeeman energies. Therefore, the degeneracy of these two modes can be lifted by applying an external magnetic field. Accordingly, we can selectively excite one of the two modes by appropriately choosing the excitation frequency. The detection of the handedness is achieved by utilizing the ISHE effect, as described in the following.

We consider a magnetic thin film consisting of perpendicularly magnetized SAF sandwiched by two HMs. The magnetizations in two FM layers in the SAF structure are coupled antiferromagnetically. The magnons in the SAFs are excited by applying a microwave magnetic field from an antenna, as schematically shown in Figs. 1b and 1c. Coherently propagating antiferromagnetic magnons in the $x$-direction emit spin currents into HMs due to the spin pumping effect. The spin current $J_s$ is converted to charge current $J_c$ by the spin-to-charge conversion due to the ISHE[30]. The resulting transverse voltage $V_{\text{ISHE}}$ in the $y$-direction is proportional to

$$F_{\text{HM1-SAF-HM2}} = -\theta_{\text{SH,HM1}} \langle \mathbf{m}_1(t) \times \dot{\mathbf{m}}_1(t) \rangle_x \\ + \theta_{\text{SH,HM2}} \langle \mathbf{m}_2(t) \times \dot{\mathbf{m}}_2(t) \rangle_x, \qquad (1)$$

where $\theta_{\text{SH,HM1(2)}}$ is the spin Hall angle of the lower (upper) HM materials, while $\langle \mathbf{m}_{1(2)}(t) \times \dot{\mathbf{m}}_{1(2)}(t) \rangle_x$ is the time average of the $x$ component of the spin polarization pumped from the lower (upper) FM layer. The external magnetic field is applied in the $xz$-plane with a tilted angle $\theta_H$ from the perpendicular axis, which tilts the precession axis of the magnetization and makes $\langle \mathbf{m}_{1,2}(t) \times \dot{\mathbf{m}}_{1,2}(t) \rangle_x$ be finite[31]. Note that the signs of two terms on the right-hand side of Eq. (1) are opposite. This is because two FM layers pump the spin current into the opposite directions ($+z$ and $-z$ for the upper and lower FM layers, respectively). The direction of the electric current generated by the ISHE depends on this pumping direction and the spin Hall angle of two HMs.

Equation (1) clarifies the relationship between the sign of the ISHE voltage and the magnon handedness for a fixed magnetic field. We notice that, for both RH and LH modes, two magnetizations in SAFs precess in the same direction, however, the oscillation amplitude is different[3], as schematically shown in the left panel of Figs. 1b and 1c. For RH (LH) mode the magnetization oriented to $+z$ ($-z$) direction is an active precession state and exhibits a larger oscillation amplitude. Accordingly, in the case of antiparallel magnetization configuration shown in Figs. 1b and 1c, the second (first) term on the right-hand side of Eq. (1) becomes the dominant contribution in the ISHE voltage for RH (LH) mode due to the large amount of $J_s$ resulting from the spin pumping effect. If $\theta_{SH,HM1}$ and $\theta_{SH,HM2}$ have the same sign such as the Pt sandwiched SAF structure, these differences in the dominant contribution change the sign of the ISHE voltage depending on the magnon handedness (RH or LH), as shown in Figs. 1b and 1c. On the other hand, if $\theta_{SH,HM1}$ and $\theta_{SH,HM2}$ have the opposite signs, both two terms on the right-hand side of Eq. (1) contribute to the ISHE voltage with the same sign. This makes it difficult to distinguish the RH and LH modes. Therefore, the use of HMs with the same sign of the spin Hall angle is a key to efficiently detect the magnon handedness.

**Sample details**

Based on the above concept, we perform the experiments. Thin films were deposited by dc magnetron sputtering on thermally oxidized Si substrate with the following structures (thickness in nm): Ta(2.0) / Pt(5.0) / [Co(0.3) / Ni(0.6)]$_{8.5}$ / Ru(0.42) / [Co(0.3) / Ni(0.6)]$_{8.5}$ / Pt(5.0) (hereafter referred to as Pt-SAF-Pt). For comparison, we also prepared perpendicularly magnetized SAF films with different combinations of HM layers, such as Ru-SAF-Pt and Pt-SAF-Ru. Note that the signs of the spin Hall angle in Ru and Pt are opposite[32], which affects the measurement as discussed above. After the deposition process, the out-of-plane magnetic hysteresis loops were measured to determine the magnetization configuration as a function of the external magnetic field (see Supplementary S1). The two different antiparallel magnetization configurations, namely head-to-head (H-H) and tail-to-tail (T-T), can be obtained depending on the field sweep direction after applying the large enough magnetic field to saturate the magnetization (see Supplementary S1). The magnetic resonance measurements were also conducted under the external magnetic field along the out-of-plane direction to quantify the values of magnetic parameters, such as the perpendicular magnetic anisotropy field and the interlayer exchange field (see Supplementary S2).

**Inverse spin Hall detection of propagating magnons**

Next, we experimentally investigate the electrical detection of the propagating magnon handedness. We fabricated the devices with the Hall bar structure and microwave antenna, as shown in Fig. 2a (see methods). The magnons are excited by applying a microwave current to the stripline antenna and the lock-in amplifier is used to detect the Hall voltage $V_{LIA}$. The detection of propagating magnon is achieved by measuring the ISHE voltage. Here the distance $d$ of the magnon propagation is 0.4 μm.

The detected signal $V_{LIA}$ under the tilted magnetic field of $\theta_H = 30°$ at the microwave frequency of $f_{rf} = 13$ GHz is shown in the inset of Fig. 2b. We notice that the hysteresis loop similar to the magnetization curve are observed. This is the contribution of the anomalous Nernst effect to the Hall voltage, because the application of microwaves causes local heating of the magnon excitation antenna, leading to the in-plane thermal gradient in the $x$-direction. This contribution can be eliminated by subtracting a linear background signal in the field range of the antiparallel magnetization configuration. The signals after background subtraction are hereafter referred to as $\Delta V_{\text{ISHE}}$. As shown in Fig. 2b, the resonance peaks in $\Delta V_{\text{ISHE}}$ spectra are observed for the upward and downward field sweep directions, corresponding to the T-T and H-H antiparallel magnetization configurations, respectively.

To clarify the relationship between the magnon handedness and the sign of ISHE resonance peak, we next measured $V_{\text{ISHE}}$ spectra at various excitation frequencies. Figures 2c and 2d show color plots of $\Delta V_{\text{ISHE}}$ spectra obtained from upward and downward field sweeps, respectively. Two precession modes are observed in the magnetic field range in which the two magnetizations are antiparallel. The resonance frequency that increases (decreases) with increasing magnetic field corresponds to the RH (LH) precession mode (see Supplementary Figs. S2a and S2b for a comparison of the magnon handedness with the magnetic resonance measurements under the out-of-plane magnetic field). The sign reversal of the resonance peaks is observed in positive and negative magnetic field regions because the in-plane component of the magnetization in the SAFs is reversed. More remarkably, the sign of the resonance peaks is also reversed depending on the handedness of the precession modes in the positive (negative) magnetic field region for T-T (H-H) antiparallel magnetization configuration, respectively. These results are consistent with our proposal mentioned in the previous section, and thus, indicate the possibility to distinguish the magnon handedness in Pt sandwiched SAFs electrically.

**Numerical simulations**

The detections of the magnon handedness demonstrated above were based on the fact that the oscillation amplitudes of the magnetization in two FM layers are different depending on the magnon handedness[3]. However, it is experimentally difficult to quantify the amounts of the pumped spin current from the upper and lower FM layers separately. Therefore, we performed numerical simulations of the Landau-Lifshitz-Gilbert equation (see methods). Figures 3a and 3b show the time average of the $x$ component of the spin polarization pumped from the lower FM layer $\langle \mathbf{m}_1(t) \times \dot{\mathbf{m}}_1(t) \rangle_x$ and upper FM layer $\langle \mathbf{m}_2(t) \times \dot{\mathbf{m}}_2(t) \rangle_x$, respectively, for T-T antiparallel magnetization configuration. As we mentioned in the previous section since the magnetization oriented to $+z$ ($-z$) direction exhibits a larger oscillation amplitude for RH (LH) mode, only the RH (LH) precession mode shows the visible peaks in Fig. 3b (Fig. 3a), respectively. Note that the sign of spin current spin polarization is positive regardless of the magnon handedness when we focus on the positive magnetic field region because the in-plane component magnetization in the active precession layer is oriented in the same direction regardless of the precession mode, as schematically shown in insets of Figs. 3a and 3b. In the case of the H-H antiparallel magnetization configuration, the relationship of magnetization orientation is opposite to that of the T-T antiparallel magnetization configuration. Therefore, only the RH (LH) precession mode shows the visible peaks in Fig. 3d (Fig. 3e), respectively. Then, the calculated $F_{\text{Pt-SAF-Pt}}$ based on Eq. (1) for both antiparallel magnetization configurations (Figs. 3c and 3f) well reproduce the observed behavior in the experiments, where the spin Hall angles are assumed to be $\theta_{\text{SH,Pt}} = 0.1$ (see Supplementary S3 for Ru-SAF-Pt and Pt-SAF-Ru structures). These calculations reveal that the spin pumping effect from the upper or lower FM is quantitatively dominant depending on the magnon handedness and the sign of the ISHE voltage for HM1-SAF-HM2 structure is determined by the spin Hall angle in HMs.

**Switching the handedness of propagating antiferromagnetic magnons**

Here, in the middle of Fig. 4, we also demonstrate the manipulation of the antiferromagnetic magnon handedness by tuning the excitation frequency $f_{\text{rf}}$ to either 13.6 GHz or 18.0 GHz under the tilted magnetic field of 140 mT for the T-T antiparallel magnetization configuration. The sign of the voltage $\Delta V_{\text{ISHE}}$ changes as the excitation frequency switches. Thus, we can switch the magnon handedness by switching the excitation frequency. The relationship between the frequency and the magnon handedness is reversed when the magnetization configuration is H-H under the tilted magnetic field of -140 mT, as shown in the bottom of Fig. 4. In summary, we have succeeded in manipulating the handedness of the propagating magnon and in electrically reading out its information, which means that we now have the technology to realize theoretically proposed novel devices using the magnon polarization degree of freedom[4–9].

## Discussion

Recently, the nonlocal transport of the magnon spin current in antiferromagnets has been observed[33–37], and an electrical detection of the magnon polarization with uniform precession has been realized by the coherent spin pumping effect in heterostructure with heavy-metal and antiferromagnetic materials[38–42]. In addition, coherent propagation of antiferromagnetic magnon has also been demonstrated using bulk antiferromagnets such as dysprosium orthoferrite (DyFeO$_3$)[43] or hematite (α-Fe$_2$O$_3$)[44–46]. However, the observation of coherently propagating magnon polarization of antiferromagnets has yet to be achieved so far. We overcome the issue by employing the SAF structure sandwiched by the Pt layer. In principle, our demonstrated method of handedness control and electrical detection of antiferromagnetic magnons is applicable to all antiferromagnets and is a major step towards harnessing the full potentials of antiferromagnetic magnons, such as the THz spin dynamics, robustness against external perturbations, and the polarization degree of freedoms.

A requirement for realizing magnon-based devices is to achieve a long length scale for transferring the information. In conventional magnon-based devices, the length scale is the magnon decay length, which is determined from the group velocity and dissipation of the propagating magnons. It is, however, difficult to estimate the characteristics of the propagating magnons, such as dispersion relation and group velocity, in the present system. This is because the ISHE detection used in the present work is not a phase-sensitive detection method, in contrast with the propagating spin wave spectroscopy using a vector network analyzer[47]. Thus, instead, we performed a micromagnetic simulation analysis (see Supplementary S4). In addition, we also performed the measurements of the propagating magnon spectra for the devices with different $d$ to evaluate the decay length of the ISHE resonance signals (see Supplementary S5). We find that the ISHE voltage arising from the magnon can be transmitted over a longer distance than the numerically calculated magnon decay length, owing to the contribution of the ISHE voltage from the magnon in the region away from the Hall bar (see Supplementary S5). Therefore, our proposed method is advantageous from the perspective of the information transfer beyond the decay length, alongside the ability to distinguish the magnon handedness.

A technical issue in our demonstration is that we need to apply a tilted magnetic field to the film plane. This is because it makes the time-averaged in-plane component of the pumped spin polarization finite and results in the generation of the dc ISHE voltage[31]. Polarization-resolved detection of magnon in a system with completely perpendicular magnetization is highly desired to enable demonstrations without the need for the magnetic field. One possible approach to achieve field-free detection is to utilize an inverse phenomenon of the spin anomalous Hall effect or the spin-orbit precession effect in ferromagnets[48–53], which can convert the out-of-plane spin polarization to a charge current. We believe this makes the experimental system simpler and encourages the further investigation of the propagating magnon handedness.

In summary, we have demonstrated a polarization-resolved detection of propagating antiferromagnetic magnons in perpendicularly magnetized SAFs by using spintronics transducers based on the ISHE. Our experimental observations and numerical simulations provide clear evidence that the magnon handedness can be identified by the sign of the ISHE voltage from the SAF structure sandwiched by the HM layers with the same sign of the spin Hall angle. In addition, we have demonstrated the switching and electrical detection of the antiferromagnetic magnon handedness by tuning the excitation frequency. These demonstrations provide a new way to detect and control the spin polarization of magnon spin currents and open up the attractive possibility of antiferromagnet-based magnon devices.

# Methods

**Device fabrication and experimental setup.**

The films were patterned into Hall-bar structures with 10-μm-wide channel width and 350-nm-wide Hall probe width using electron beam lithography and Ar ion milling process, and then deposited $SiO_2$ (50 nm) for electrical isolation of the microwave antenna. Subsequently, we fabricated contact pads for the Hall probes and 500-nm-wide single-stripline microwave antenna using electron beam lithography and lift-off process. The microwave antennas were made of Ti (5 nm) / Au (80 nm) with the distance $d$ between the edges of the antenna and the Hall probe. In the measurement setup (Fig. 2a), the microwave antenna was connected to a signal generator (Anritsu; 68347C), and a pulse-modulated microwave was applied, whereas the Hall probe was connected to a lock-in amplifier (Stanford Research; SR830) to detect the transmission of propagating magnons along the $x$-axis. The applied microwave power is 10 mW (10 dBm). To improve the signal-to-noise ratio, the measurements were repeated 4 times and averaged spectra are plotted in the figure.

**Numerical simulations.**

For numerical study, we solved the Landau-Lifshitz-Gilbert equation with the unit vector of magnetization of lower and upper FM layers, $\mathbf{m}_1$ and $\mathbf{m}_2$, given by

$$\dot{\mathbf{m}}_i = -\mu_0 \gamma \mathbf{m}_i \times \mathbf{H}_i + \alpha \mathbf{m}_i \times \dot{\mathbf{m}}_i, \tag{2}$$

where $\mu_0$, $\gamma$, and $\alpha$ are the vacuum permeability, the gyromagnetic ratio, and the Gilbert damping constant, respectively. The effective magnetic field $\mathbf{H}_i$ acting on the magnetization is given by

$$\mathbf{H}_i = \mathbf{H}_{\text{ext}} + H_{k,i}^{\text{eff}} m_{z,i} \mathbf{e}_z - H_E \mathbf{m}_j \quad (i \neq j), \tag{3}$$

where $\mathbf{H}_{\text{ext}}$ is the external magnetic field, $H_{k,i}^{\text{eff}}$ is the effective perpendicular magnetic anisotropy fields, and $H_E = -J_{\text{ex}}/(\mu_0 M_s t_{\text{FM}})$ is the interlayer exchange field with the interlayer exchange coupling energy per unit area $J_{\text{ex}}$ ($J_{\text{ex}} < 0$ indicates the antiferromagnetic coupling), the saturation magnetization $M_s$, and the thickness of FM layer $t_{\text{FM}}$. Here, we use a uniform precession motion for simplicity, which is sufficient to distinguish the contributions of the upper and lower HM layers. Thus, the dynamic dipolar interaction[54] is excluded from the effective magnetic field. Equation (2) was numerically solved by substituting Eq. (3) by the fourth-order Runge-Kutta method with the time interval $\Delta t = 0.1$ ps under the application of alternating magnetic field $h_{\text{rf}} \cos(2\pi f t)$ along the $x$-axis, where $\mu_0 h_{\text{rf}} = 1.0$ mT.

**Data availability**

The data that support the findings of this work are available from the corresponding authors upon reasonable request.

**Acknowledgments**

This work was partly supported by JSPS KAKENHI Grant Numbers JP22H01936, JP20H05665, and JP23KK0093; a research granted from The Murata Science Foundation; MEXT Initiative to Establish Next-generation Novel Integrated Circuits Centers (X-NICS) Grant Number JPJ011438; and the Collaborative Research Program of the Institute for Chemical Research, Kyoto University.

**Author contributions**

Y.S. and T.O. conceived the project. Y.S. deposited the multilayers and fabricated the devices. Y.S. performed the measurements and data analysis. Y.S. and T.T. performed the simulation. All authors contributed jointly to the interpretation of the results. Y.S., T.T., and T.O. wrote the manuscript with the assistance of the other authors.

**Competing interests**

The authors declare no competing financial interests.

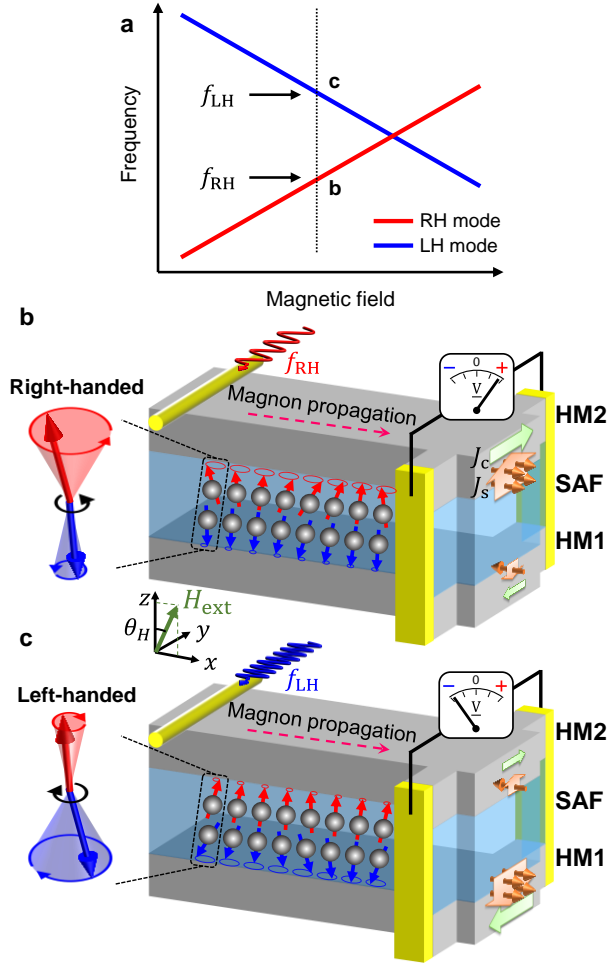

**Fig. 1 Handedness manipulation of propagating antiferromagnetic magnon**. **a**, Magnetic resonance frequency for right-handed (RH) and left-handed (LH) precession modes in colinear antiferromagnets as a function of the magnetic field. **b,c,** Left panels, illustrations of magnon eigenmode with RH and LH precession modes. The handedness of propagating antiferromagnetic magnons in perpendicularly magnetized synthetic antiferromagnet (SAF) can be manipulated by tuning the excitation microwave frequency to either $f_{RH}$ or $f_{LH}$ at a fixed magnetic field, as indicated by the black dotted line in (**a**), which corresponds to the condition where the degeneracy of two magnon modes is lifted. The handedness-dependent spin current $J_s$ is emitted through the spin pumping effect into heavy metal (HM) layers and its $x$-polarized component shown in the figure is converted to charge current $J_c$ through the inverse spin Hall effect.

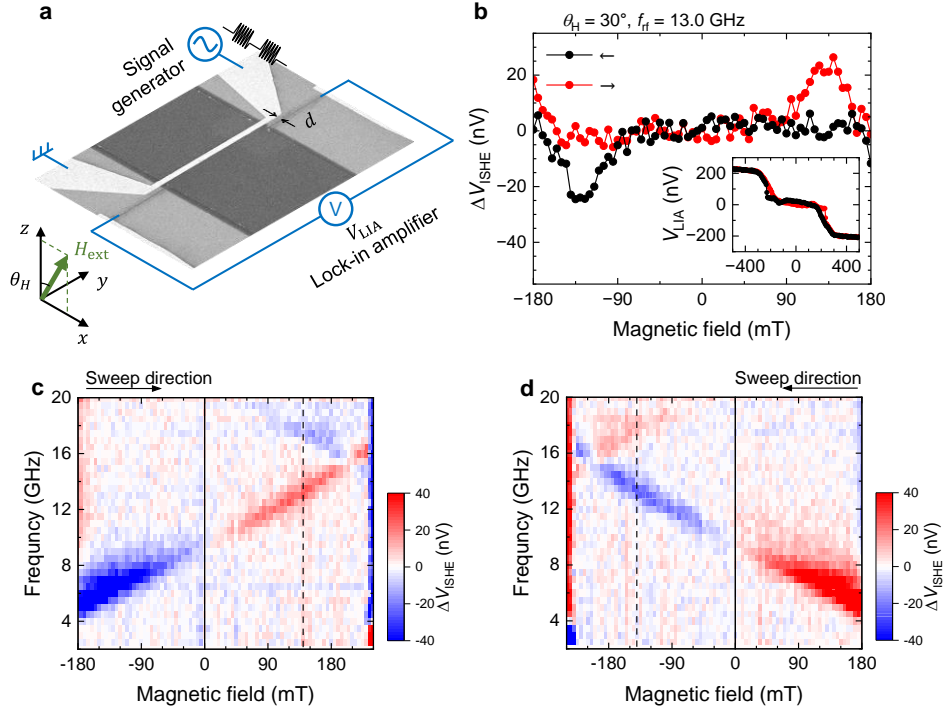

**Fig. 2 Inverse spin Hall detection of propagating magnons in Pt-SAF-Pt**. **a,** Schematic of the measurement setup and scanning electron microscope image of the devices. **b,** The signal after background subtraction $\Delta V_{ISHE}$ under the tilted magnetic field of $\theta_H = 30°$ at the microwave frequency of 13 GHz. The black and red curves indicate the spectra measured for the downward and upward field sweep direction, respectively. The inset shows the detected signal $V_{LIA}$ under the tilted magnetic field. **c,d,** $\Delta V_{ISHE}$ spectra under the tilted magnetic field $\theta_H = 30°$ for T-T antiparallel magnetization configuration (**c**) and H-H antiparallel magnetization configuration (**d**). The dashed lines in (**c,d**) cprrespond to the magnetic field for the experiments performed in Fig. 4.

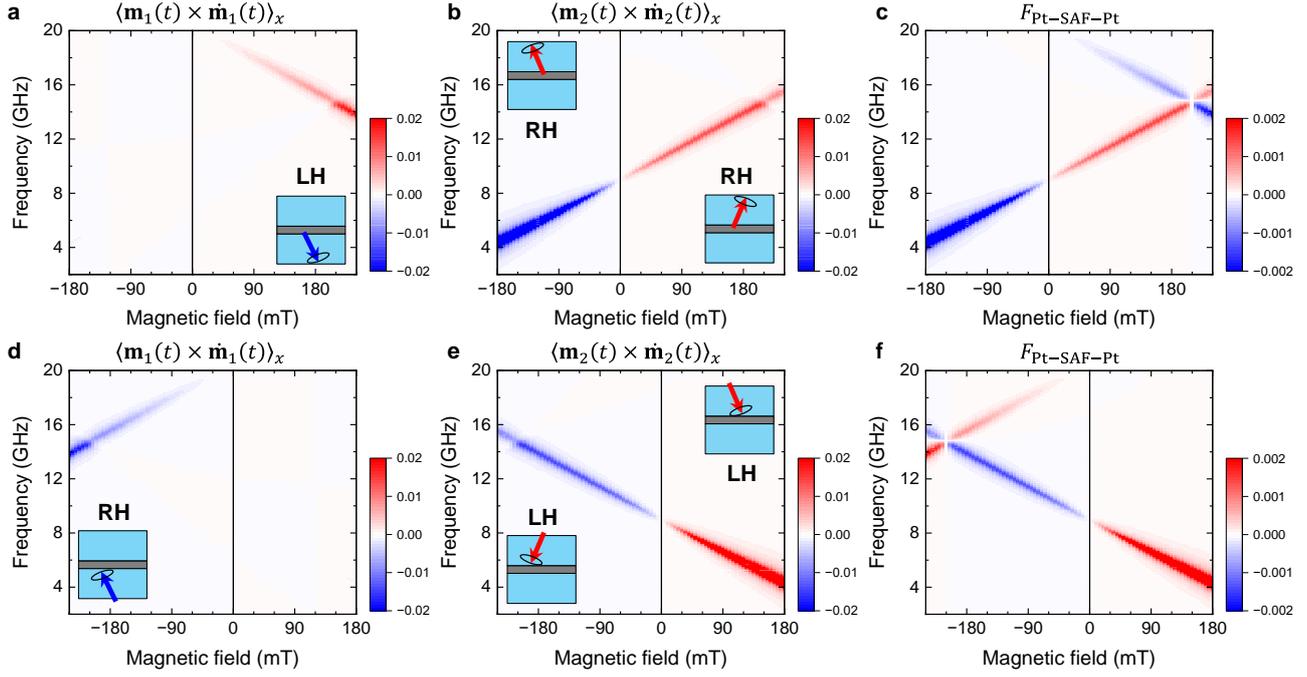

**Fig. 3 Calculated spin polarization and expected ISHE voltage arising from the propagating antiferromagnetic magnons in Pt-SAF-Pt. a-f,** Color plots of calculated $\langle \mathbf{m}_1(t) \times \dot{\mathbf{m}}_1(t) \rangle_x$ (**a,d**), $\langle \mathbf{m}_2(t) \times \dot{\mathbf{m}}_2(t) \rangle_x$ (**b,e**), and $F_{\text{Pt-SAF-Pt}}$ based on Eq. (1) (**c,f**) for T-T antiparallel magnetization configuration (**a-c**) and H-H antiparallel magnetization configuration (**d-f**). The insets show the illustration of SAFs with active precession magnetization state for RH and LH modes.

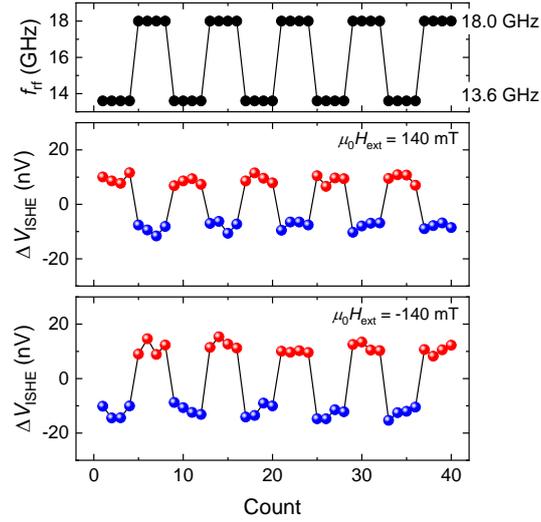

**Fig. 4 Switching the handedness of propagating antiferromagnetic magnons**. As shown in the top panel, the excitation microwave frequency $f_{rf}$ is tuned to either 13.6 GHz or 18.0 GHz, and $\Delta V_{ISHE}$ is detected in T-T antiparallel magnetization configuration at $\mu_0 H_{ext}$ = 140 mT (middle panel) and in H-H antiparallel magnetization configuration at $\mu_0 H_{ext}$ = -140 mT (bottom panel), with the external magnetic field tilted at $\theta_H$ = 30°.